\documentclass{iopconfser}
\usepackage{graphicx}
\usepackage{enumitem}

\begin{document}

\title{Evaluating Application Characteristics for GPU Portability Layer Selection}

\author{
Mohammad Atif$^{1}$, 
Meghna Bhattacharya$^{2}$,
Mark Dewing$^{2}$,
Zhihua Dong$^{1}$,
Julien Esseiva$^{3}$,
Oliver Gutsche$^{2}$,
Matti Kortelainen$^{2}$,
Ka Hei Martin Kwok$^{2}$,
Charles Leggett$^{3}$,
Meifeng Lin$^{1}$,
Aleksei Strelchenko$^{2}$,
Vakhang Tsulaia$^{3}$,
Brett Viren$^{1}$, 
Tianle Wang$^{1}$,
Haiwang Yu$^{1}$,
}

\affil{$^1$ Brookhaven National Laboratory, Upton, NY 11973, USA }
\affil{$^2$ Fermi National Accelerator Laboratory, Batavia, IL 60510, USA }
\affil{$^3$ Lawrence Berkeley National Laboratory, Berkeley, CA 94720, USA }
\affil{$^4$ Argonne National Laboratory, Lemont, IL 60439, USA }

\email{cgleggett@lbl.gov}

\begin{abstract}
GPUs have become the dominant source of computing power for high performance computing and are increasingly being used across the High Energy Physics computing landscape for a wide variety of tasks. Though NVIDIA is currently the main provider of GPUs, AMD and Intel are rapidly increasing their market share. As a result, programming using a vendor-specific language such as CUDA can significantly reduce deployment choices. There are a number of portability layers such as Kokkos, Alpaka, SYCL, OpenMP and std::par that permit execution on a broad range of GPU and CPU architectures, significantly increasing the flexibility of application programmers. However, each of these portability layers has its own characteristics, performing better at some tasks and worse at others, or placing limitations on aspects of the application. In this presentation, we report on a study of application and kernel characteristics that can influence the choice of a portability layer and show how each layer handles these characteristics. We have analyzed representative heterogeneous applications from CMS (patatrack and p2r), DUNE (Wire-Cell Toolkit), and ATLAS (FastCaloSim) to identify key application characteristics that have different behaviors for the various portability technologies. Using these results, developers can make more informed decisions on which GPU portability technology is best suited to their application.
\end{abstract}

\section{Introduction}

\begin{figure}[h]
\begin{center}
\includegraphics[width=0.8\textwidth]{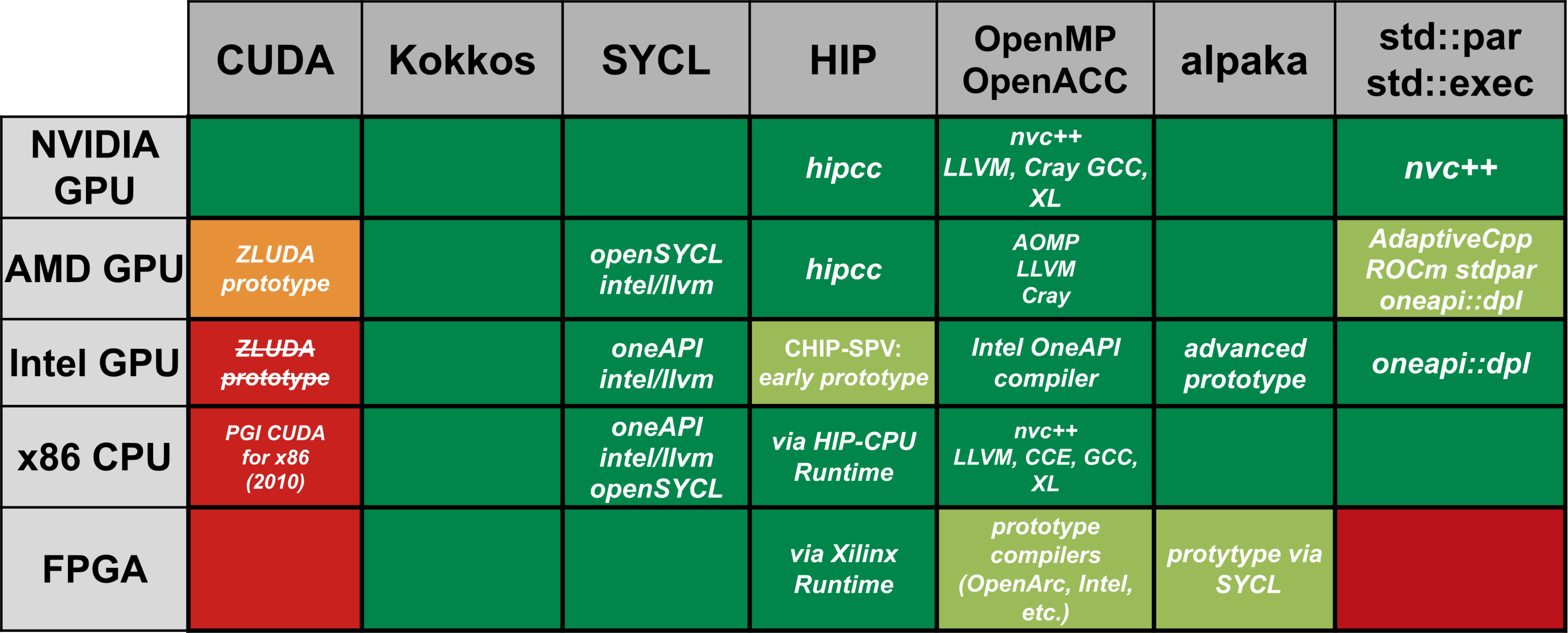}
\caption{\label{PL_matrix}Hardware support of portability layers. Dark green indicates full support, light green indicates partial support or that the project is still under development, orange is a proof of concept with an uncertain future, and red indicates no support (though there may have been support in the past).}
\end{center}
\end{figure}

Over the past several years, great progress has been made by various vendors to support a broad range of GPU and other computational accelerator backends in a portable manner. Kokkos~\cite{edwards:2014, trott:2022}, SYCL~\cite{reyes2016sycl}, HIP~\cite{kwack2021evaluation}, OpenMP~\cite{bak2022openmp}, alpaka~\cite{mathes:2017} and std::par~\cite{lin2024preliminary} can all target NVIDIA, AMD and Intel GPUs, as well as multicore CPUs and FPGAs with various degrees of completeness (see Figure~\ref{PL_matrix}). However, just because an architecture can be targeted, doesn't mean that all portability layers are equivalent. When developers are faced with the task of writing or converting code to run on GPUs, an informed decision needs to be made as to which technology to use, and which is best suited for their application. Understanding how different portability technologies handle various use cases, patterns and code structures is very useful in predicting how an application will perform before any code is written. 

The High Energy Physics Center for Computational Excellence (HEP-CCE) has spent several years porting a number of representative HEP testbeds from serial C++ code to a variety of portability technologies. We have used these testbeds to identify several code characteristics which can significantly affect an application's performance with specific portability technologies.

\section{HEP-CCE Testbeds}
The testbeds that the HEP-CCE have chosen are from a number of HEP experiments, including ATLAS, CMS and DUNE. These testbeds are relatively small code bases that can be built and executed outside of the experiments’ full frameworks in order to simplify the development process. From ATLAS, we have FastCaloSim\cite{fcs}\cite{fcs_gpu}\cite{fcs_chep}, a standalone, parametrized simulation of the ATLAS Liquid Argon Calorimeter. From CMS we chose Patatrack\cite{patatrack}, a standalone version of the heterogeneous pixel reconstruction which spans 40+ kernels, implements multithreaded concurrency, and utilizes memory pools, as well as P2R\cite{p2r}, which performs the propagate-to-radius track reconstruction in a single kernel. From DUNE we have extracted the Liquid Argon Time Projection Chamber simulation module\cite{wct}\cite{lin2023portable}, which is comprised of three kernels that perform a rasterization, scatter-add, and FFT convolution. We have also used the GPU ports of Sherpa (now called Pepper) and Madgraph, which are leading order event generators that make use of auto-generated code.

\section{Kernel Runtime and Launch Latency}

Launching a kernel from a host device onto a GPU is not an instantaneous process. Even under the best of conditions, this can take from a few to many tens of microseconds. This launch latency is dependent on many factors, such as the GPU architecture (in general AMD GPUs have much higher launch latency than NVIDIA), the CPU architecture, the clock speed of the CPU and that of the bus between CPU and GPU, as well as GPU driver version and size of the parameter list passed to the kernel.

Some portability layers can significantly increase this launch latency. Kokkos can add tens of microseconds, and even many tens of microseconds when executing on AMD GPUs, which in general have much larger launch latencies than NVIDIA or Intel GPUs. Large additional penalties were not observed for SYCL, HIP, OpenMP or Alpaka. We also observed that calling \verb|rocRAND| (the hardware native random number generator from AMD) on AMD GPUs resulted in very large launch latencies, sometimes exceeding 1s, on the first call.

If the runtime of a kernel, or a group of sequential kernels, is short and on the order of the launch latency, the performance of the application can suffer greatly. While there are techniques whereby the latency can be hidden, such as with asynchronous kernel launches where work continues on the CPU while the GPU kernel executes, this is not always possible. In these situations, where the kernel runtime is very short, it is best to avoid portability layers that exacerbate the problem.

\section{Concurrency and Thread Pools}

In HEP codes, GPU applications are rarely standalone. Rather they exist within the context of a larger application framework, which often use concurrency mechanisms such as multi-threading or multi-processing to improve performance or make better use of resources. This means that not only can the GPU kernels be launched, sometimes concurrently, from different threads or processes, but that the portability layer itself also needs to support the larger concurrency mechanism.

We found that Kokkos has significant incompatibilities with concurrency and thread pools such as Intel's Threading Building Blocks (TBB). The serial backend of Kokkos implements a lock that serializes concurrent calls. The threads backend explicitly forbids calls from external threads. This results in very poor performance for multi-threaded applications that attempt to concurrently launch kernels. Furthermore, the concurrent launch of GPU kernels is only achievable with CUDA and HIP backends, using architecture specific APIs that limit portability. (Kokkos is however developing a prototype experimental feature of "partitioned execution spaces" which may address this in the near future.)

We also observed that the SYCL implementations were inconsistent in their ability to launch GPU kernels. Some implementations on the same architecture serialized concurrent kernel calls from different threads. This is often very dependent not only on the compiler manufacturer, but also on the version of the compiler itself, demonstrating how the SYCL implementations continue to evolve rapidly. Furthermore, some architectures, such as AMD, do not appear to have any concurrent SYCL solution.

\section{External Library and Compiler Compatibility}

Many HEP applications use a variety of external libraries such as ROOT or Eigen (a commonly used C++ template library for linear algebra). These can have problematic interactions with some portability layers. For example, many versions of Eigen are incompatible with the NVIDIA CUDA compiler \verb|nvcc|. Since portability layers such as Kokkos and Alpaka use \verb|nvcc| behind the scenes to compile code for NVIDIA devices, this will also affect them. Eigen is also currently incompatible with OpenMP offload.

ROOT cannot yet be fully compiled with the NVIDIA C++ compiler \verb|nvc++|, though NVIDIA compiler developers are aware of the issues, and are working toward a solution. As a result, applications that use ROOT cannot be entirely compiled with \verb|nvc++|. It is however possible to compile part of an application with \verb|g++|, and part with \verb|nvc++|, and link them together, as they are link compatible, however care must be taken to not expose the ROOT headers to \verb|nvc++|, and the final linking must be performed by \verb|nvc++|. In this situation, any data that will be offloaded to a GPU must be allocated by \verb|nvc++|, which may require copying data if the structure was previously allocated by \verb|g++|. Furthermore, using two C++ compilers within a single project can also enormously complicate the build rules.

Similarly, care must be taken when building a project that supports multiple portability layers, as they can often be incompatible. For example, some primary Kokkos headers cannot be exposed directly to \verb|nvcc| (Kokkos requires the use of the \verb|nvcc-wrapper| compiler wrapper), so for a project that will support both Kokkos and CUDA from the same sources, careful use of \verb|#ifdef| or file separation and build rules must be used.

\section{Data Structure Complexity and Memory Transfers}

When event data models for a number of large HEP experiments were being designed, it was often very common to make use of complex, object-oriented data structures. Unfortunately, this style of data structures maps very poorly onto GPUs, which prefer flat layouts or simple Structure of Arrays. Polymorphic object types are also poorly supported in GPU languages such as CUDA, HIP and SYCL.

Portability layers like Kokkos, SYCL and Alpaka offer additional support for memory constructs such as \verb|Kokkos::Views|, and \verb|sycl::buffers| to enable seamless portability across GPU and CPU architectures. However, this portability come with a price - that of increased overheads for allocations and data transfers, especially when using many small objects. By default, Kokkos will perform a memory initialization of all new \verb|Kokkos::Views|, which can significantly impact performance. Depending on the code, this may not be necessary and can be avoided by using \verb|Kokkos::ViewAllocateWithoutInitializing|.

Alternatively, memory on the GPU can be allocated using device native calls, such as \verb|cudaMalloc|, and the pointers can be wrapped by portable structures such as \verb|Kokkos::Views|. While this mechanism may be more performant than the creation and device allocation of the portable structure, it comes with a loss of portability, as the code must determine the desired hardware backend at compile time and call the appropriate device native allocator.

We have also observed that automatic memory transfers using Unified Shared Memory (at least for discrete GPUs) are invariably slower than explicit memory transfers, and may not occur at the optimal time. Note that std::par can only perform USM transfers, which is done by instrumenting constructors at compile time and triggering transfers on page faults during data accesses.

It should also be noted that none of the APIs can gracefully represent multidimensional vectors of varying sizes (ie jagged arrays), which is a structure that tends to be heavily used in HEP codes. Instead jagged vectors must be crafted by hand, reduced in dimension, or padded to regular dimensions.

\section{RNGs, FFTs, Atomics and Portability}

NVIDIA, AMD and Intel have all provided GPU native implementations of random number generators (cuRAND, rocRAND, and oneMKL), and Fast Fourier Transforms, which are tuned to their architectures. Unfortunately their APIs are not portable. In general, portability layers such as Kokkos and Alpaka do not offer APIs that target the native functions though Kokkos does provide its own implementations of RNGs and FFTs, which are consistent across architectures. While it is in general possible to call the device functions from the portability layer, eg calling \verb|cuRAND| from Kokkos code when compiling for NVIDIA GPUs, this does not make for portable code. The HEP-CCE group has helped develop hardware native backends to oneMKL which permit SYCL to call the device native implementations through a common interface on Intel, NVIDIA and AMD GPUs \cite{RNG}. This feature has not yet been extended to FFTs. We have also developed a header only library which wraps the device native RNGs with a uniform API and can target NVIDIA and AMD GPUs as well as CPUs\cite{RNG2}. 

It is often necessary to use \verb|atomics| in HEP codes, though these operations map poorly onto GPU hardware. Some older GPU hardware do not natively support atomic operations for all \verb|int|, \verb|float| and \verb|double| types, requiring software emulation and resulting in very poor performance. Alternatives to atomic operations often involve significant and onerous restructuring of code for marginal performance benefit. While all portability layers now support some form of atomic operations, such as the \verb|Kokoss::atomic_fetch_add()| or \verb|sycl::atomic_ref|, this was not always the case, and performance can vary wildly. For example using \verb|atomic<int>| on NVIDIA hardware using OpenMP and \verb|clang13| is \verb|35x| slower than an equivalent CUDA \verb|atomicAdd| call. We see much more comparable performance with more recent versions of Clang. Integer performance of atomics is also consistently worse than that of floats and doubles. Furthermore, \verb|nvc++| requires \verb|C++17| to support \verb|atomic<int>|, and \verb|C++20| for \verb|atomic<float>|.

\section{Compilation Time}

For small projects, such as those used in this analysis, compilation time is rarely an issue. However, larger code bases can have very significant compilation times, such as for example the ATLAS reconstruction code which can take several hours for a full build on modern hardware. There can be serious impacts on the developer cycle if a portability layer significantly increases it. 

We have observed that in general, \verb|nvc++| is \verb|2x| to \verb|3x| slower than \verb|g++|, and Kokkos adds a \verb|10%| to \verb|20%| overhead depending on code complexity and the use of templates. MadGraph, with its use of heavily templated, autogenerated code, had exceedingly long compilation times when using the \verb|icpx| backend of SYCL for Intel GPUs, which were sometimes more than \verb|1000x| slower than other backends for large kernels. This issue is currently being addressed by Intel compiler developers.

\section{Runtime Provisioning}

When using a portability layer, provisioning a project for execution on a specific or range of architectures is not entirely trivial. Kokkos needs to be built with very specific hardware architecture identifiers. One can only select a single set of serial, host parallel and device parallel combinations, and the architectures, especially for GPUs, need to be very specific, as Kokkos attempts to perform low level optimizations for specific hardware devices. For example, if code is built for an NVIDIA V100 GPU, it will not be able to run on an NVIDIA A100 GPU. Furthermore, if support for shared libraries is desired, Kokkos must be built with specific flags, and then care must be taken with code organisation so that offloaded symbols in one GPU kernel cannot be split between multiple compilation units. This can result in a very large matrix of build configurations which must all be compiled separately and then distributed, leading to very challenging binary distribution mechanisms when the runtime architecture of remote sites is not precisely known \textit{a-priori}. 

For Alpaka device code, each backend needs to be compiled with the backend specified compiler, such as \verb|nvcc|, \verb|hipcc| or \verb|icpx|. While binaries from all the backends can be loaded into the same process at runtime, user code must guarantee unique symbols between the backends, and the backend technologies must not conflict.

OneAPI has recently begun to support multi-platform binaries, which must be identified at compile time. Only one device per Intel/AMD/NVIDIA architecture may be specified, which again causes issues for code distribution if the runtime architecture is not explicitly known beforehand. Other SYCL implementations do not currently support multi-platform binaries.

For some OpenMP implementations, such as LLVM Clang, multiple architectures can be listed at compile time via the \verb|--offload-arch| flag. At runtime, the backend can be dynamically selected by querying the system to discover what specific hardware is available.

\section{Conclusions}

In general, there is no single GPU portability solution that is optimal for all situations. What works well for one project may prove sub-optimal for another. When selecting a GPU portability layer, a careful analysis of an application's characteristics is important to achieve the best match to ensure compatibility, performance and portability. Selecting a portability layer before fully understanding the nature of the code, its externals, and the full runtime environment will result in significant issues for both provisioning and performance. 

It is very important to note that Kokkos, Alpaka, SYCL, OpenMP and std::par continue to evolve quite rapidly, as do the native compilers such as \verb|clang|, \verb|nvcc|, \verb|nvc++|, \verb|hipcc|, and \verb|icpx| which they use behind the scenes. Careful and frequent monitoring of the technologies and architectures are necessary to evaluate the match between a code base and portability layer. Fortunately most of the changes seem to be in a positive direction, and we are seeing the convergence of increasing feature support and optimized performance in all the portability layers.

It's also worth noting that the emerging C++ standards for concurrency and offloading such as the \verb|P2300|\cite{p2300} proposal for asynchronous execution in C++ may simplify choices in the not too distant future, though it will still be several years before they are supported by a broad range of hardware.

\bibliography{refs}
\bibliographystyle{iopart-num}

\end{document}